\documentclass[fleqn,10pt]{wlscirep}
\usepackage[utf8]{inputenc}
\usepackage[T1]{fontenc}

\usepackage[
  separate-uncertainty = true,
  multi-part-units = repeat
]{siunitx}

\usepackage{float}
\usepackage{cite}
\usepackage{amsmath,amssymb,amsfonts}
\usepackage{algorithmic}
\usepackage{graphicx}
\usepackage{tabularx, booktabs, makecell, caption}
\usepackage{siunitx}
\def\arrvline{\hfil\kern\arraycolsep\vline\kern-\arraycolsep\hfilneg}
\usepackage{textcomp}

\begin{document}
\title{Swap-Free Fat-Water Separation in Dixon MRI using Conditional Generative Adversarial Networks}

\author[1,+]{Nicolas Basty}
\author[1]{Marjola Thanaj}
\author[2]{Madeleine Cule}
\author[2]{Elena P. Sorokin}
\author[2]{Yi Liu}
\author[1]{Jimmy D. Bell}
\author[1]{E. Louise Thomas}
\author[1]{Brandon Whitcher}

\affil[1]{Research Centre for Optimal Health, School of Life Sciences, University of Westminster, London, UK}
\affil[2]{Calico Life Sciences LLC, South San Francisco, California, USA}

\affil[+]{email: n.basty@westminster.ac.uk}
\date{}

\begin{abstract} 
Dixon~MRI is widely used for body composition studies. Current processing methods associated with large whole-body volumes are time intensive and prone to artifacts during fat-water separation performed on the scanner, making the data difficult to analyse. The most common artifact are fat-water swaps, where the labels are inverted at the voxel level. It is common for researchers to discard swapped data (generally around 10\%), which can be wasteful and lead to unintended biases. The UK Biobank is acquiring Dixon MRI for over 100,000 participants, and thousands of swaps will occur. If those go undetected, errors will propagate into processes such as abdominal organ segmentation and dilute the results in population-based analyses. There is a clear need for a fast and robust method to accurately separate fat and water channels. In this work we propose such a method based on style transfer using a conditional generative adversarial network. We also introduce a new Dixon loss function for the generator model. Using data from the UK~Biobank Dixon MRI, our model is able to predict highly accurate fat and water channels that are free from artifacts. We show that the model separates fat and water channels using either  single input (in-phase) or dual input (in-phase and opposed-phase), with the latter producing improved results. Our proposed method enables faster and more accurate downstream analysis of body composition from Dixon~MRI in population studies by eliminating the need for visual inspection or discarding data due to fat-water swaps.

\end{abstract}
\maketitle

\section{Introduction}\label{sec:introduction}

The UK Biobank abdominal imaging protocol produces several MRI datasets that focus on basic structural and metabolic measurements in the thorax, abdomen and pelvis~\cite{littlejohns2020uk}.  The primary dataset in this protocol is a series of six acquisitions covering the neck-to-knee area of the body using a two-point Dixon method~\cite{dixon1984simple}, where two 3D T1-weighted volumes are acquired with modified spin-echo sequences resulting in voxel signal intensities that depend on the difference between magnetization in fat and water. The presence of field inhomogeneities in data acquisition can cause errors in fat-water separation performed on the scanner. When a voxel is incorrectly labelled during the separation process performed inside the scanner, this is known as a fat-water swap and usually occurs in contiguous sets of voxels associated with a specific tissue type, assuming a region-growing algorithm has been used. In most studies, including the UK~Biobank, the phase information is not provided to the end user making a full reconstruction impossible.  


Over the last decade, MRI has become the gold standard for body composition, particularly when measuring adipose tissue, liver and pancreatic fat content. Some of these measurements have had an enormous impact on our understanding of metabolic conditions such as type-2 diabetes and nonalcoholic fatty liver disease~\cite{thomas2013whole}. In addition to these measurements, the data from UK~Biobank abdominal protocol covers multiple tissues and organs such as muscles, abdominal organs, bones, adipose tissue, etc, with the potential for a myriad of clinically relevant variables. It is therefore somewhat surprising that to date there have been relatively few publications arising from the abdominal protocol (20), compared to those arising from the brain (168) and cardiac (89) protocols in the UK~Biobank. These results come from individual PubMed searches using the keywords: ``UK~Biobank'', ``MRI'' and either ``Abdominal'', ``Brain'' or ``Cardiac''. This is also reflected in the paucity of abdominal image-derived phenotypes that have been returned to the UK Biobank, which in turn reflects the lack of robust automated methods for a comprehensive analysis of the abdominal MR images, especially the Dixon sequence.

Previous work on Dixon~MRI from the UK~Biobank has identified via visual inspection approximately 4\% of the participants with at least one fat-water swap in the first 40,000 scans~\cite{langner2020large, langner2020kidney} or as we have previously reported via neural-network based techniques \cite{liu2021genetic}. Our previous method was part of an extensive image processing pipeline and relied on six individual models trained specifically for each of the Dixon series as a whole, except the final two series where we split the volume in half to account for the fact that the legs are likely to be swapped individually as they are unconnected structures. Given the fat and the water channels, the model assigns a label to both channels, a swap is identified if the labels do not match and the data are swapped back to the correct label. Using this technique, we were able to correct the majority of the swaps, though these models were not designed to identify more complex swaps, including those that did not affect the entire series or, in the case of the legs, one half of the series.

\begin{figure*}[!ht]
    \centering
    \includegraphics[width=\textwidth]{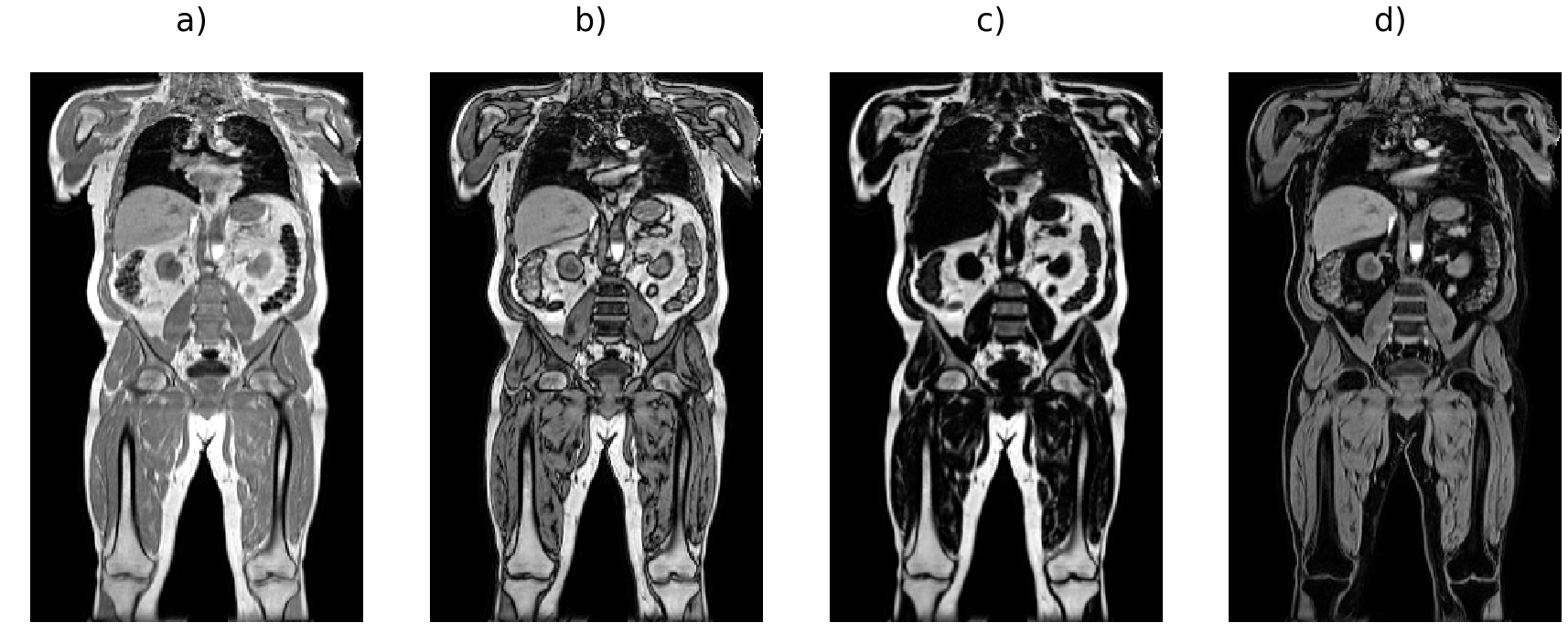}
    \includegraphics[width=\textwidth]{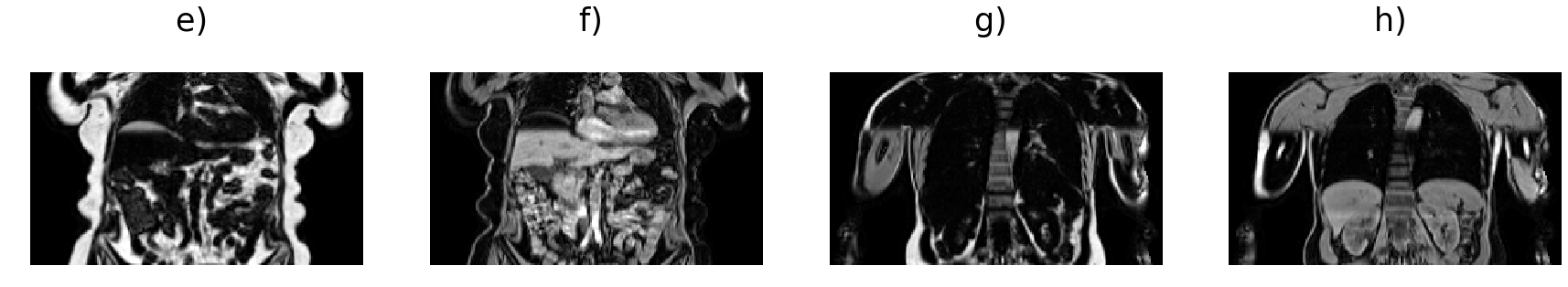}
    \includegraphics[width=\textwidth]{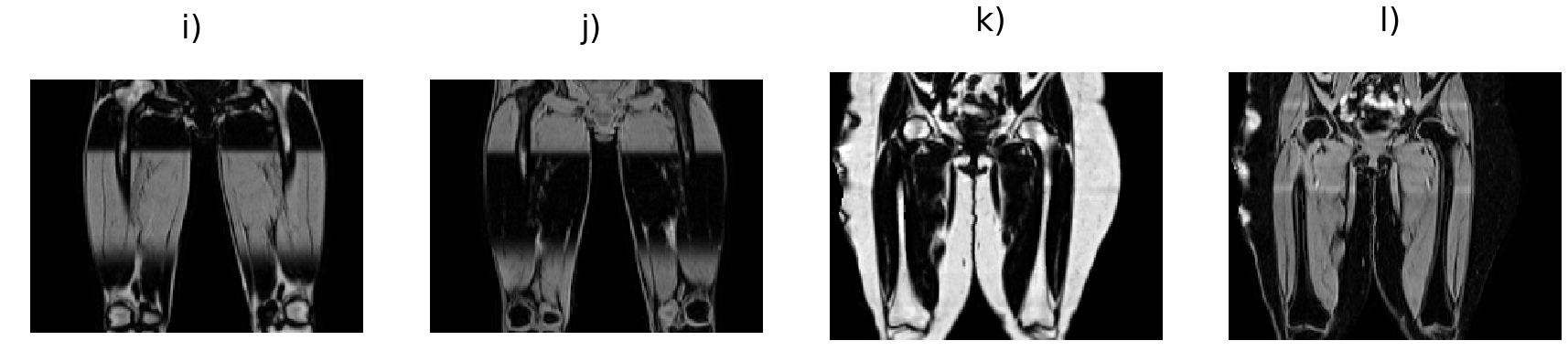}
    \caption{Example Dixon MRI data provided by UK Biobank as is: assembled in-phase (a), opposed-phase (b), fat (c) and water (d) channels. Single series of fat and water channels with a partial swap at the top of the liver (e,f), a swap in a portion of the left arm and torso (g,h), a swap in both legs (i,j), and  localised swaps due to field inhomogeneities at the boundary of the field of view (k,l).}
    \label{fig:data_and_swaps}
\end{figure*}

There are indeed a variety of more complex fat-water swaps in the UK~Biobank Dixon MRI datasets, including partial swaps that cover only a fraction of the volume or swaps related to the multiple (six) series acquired in the UK~Biobank abdominal protocol. For example, the top of the liver may be swapped when it appears in the second series and is isolated from all other tissue by the lungs (Fig.~\ref{fig:data_and_swaps}~e-f), or one of the abdominal muscles may be swapped when isolated from all other tissue by internal fat. Localised fat-water swaps may also occur due to field inhomogeneities at the boundary of the field of view (Fig.~\ref{fig:data_and_swaps}~k-l). Swaps occur more frequently in subjects of extreme sizes, therefore not being able to quantify or even completely discarding these subjects may introduce bias in population studies. The UK~Biobank is currently acquiring images for 100,000 participants, and plans to scan as many of them as possible for a second time as well as performing a separate COVID-19 study. This means that substantially more than 100,000 scans will be performed, thus we estimate that the images from close to 10,000 participants could be affected. It is essential that we develop methodologies that will ensure minimal data waste and that all analyses resulting from this impressive collection will reach their full potential. 

While some reported analyses of Dixon MRI choose to visually identify and discard data with fat-water swaps, reducing the overall number of subjects in the study by generally around 10\%~\cite{andersson2019mri,henninger20173d,henninger2021performance,ladefoged2014impact,langner2020large,langner2020kidney}, others have applied correction in post-processing, with improvements in separation techniques being an active area of research~\cite{ma2008dixon,hu2012ismrm}. Indeed, multi-point Dixon sequences ($>2$ echoes) have been developed partly to overcome this issue. However, they come with the cost of longer acquisition times and may not be practical in protocols where speed is the essence, so the two-point Dixon method will continue to be heavily utilised. In the case of the UK Biobank, the implementation of the two-point Dixon is a result of a trade-off between acquisition time, image quality and anatomical coverage. 

Few postprocessing methodologies have been proposed that both detect and correct fat-water swaps. Glocker et~al.~\cite{glocker2016correction} proposed a method for automated swap correction using machine learning, where fat-water swaps are labeled using all four channels as input to identify the locations in whole-body volumes. Correcting the fat-water swaps was performed by inverting the voxels (fat-to-water and water-to-fat) that were identified as swapped. Our previous method for swap detection and correction on a series-by-series basis also falls into this category~\cite{liu2021genetic}.

A variety of methods for improving fat-water separation have been developed, including methods based on region growing~\cite{liu2016two,cheng2017fat,yu2005field,samsonov2019resolving,triay2019magnitude}, spatial smoothing~\cite{reeder2004multicoil,andersson2018water}, graph cuts~\cite{berglund2017multi,baselice2017modified}, patch-based methods \cite{zhao2016identification} and the projected power method \cite{zhang2017resolving}. The IDEAL (iterative decomposition of water and fat with echo asymmetry and least squares estimation) algorithm performs separation via estimating errors in the field map~\cite{reeder2004multicoil}.

Recently, deep-learning methods based on convolutional neural networks (CNNs) have been applied to separate the in-phase and opposed-phase data from a two-point Dixon acquisition into fat and water channels in 2D: fat-water separation and parameter mapping in cardiac MRI~\cite{goldfarb2019water} and fat-water separation in whole-body Dixon MRI by predicting the full volume slice-by-slice with the real and imaginary parts of the first echo time~\cite{andersson2018single,andersson2019separation}. Two-echo input data was used for fat-water separation in~\cite{zhang2018dual}. In \cite{cole2021analysis}, authors showed that using complex data outperformed real data on its own for phase-based applications and reconstructions of 2D data for three different datasets. The authors in~\cite{shih2021deep} propose using CNNs for parameter mapping and uncertainty estimation for fat quantification. Another recent effort for fat-water separation in 2D data that simultaneously estimates R2* and field decay has been proposed by~\cite{jafari2021deep}. The authors also propose unsupervised training without using labels to exploit the underlying physical model, and show good agreement between models that used labels and those that did not. A similar self-supervised learning strategy is applied to reconstruction of R2* data in~\cite{torop2020deep}. All the above methods were based or closely related to the U-Net architecture~\cite{ronneberger2015u} for 2D data. A bi-directional convolutional residual network has been shown to outperform the common U-Net in multi-echo gradient recalled echo data, where the results improve when increasing the number of echo times~\cite{liu2020robust}. In~\cite{cho2019robust}, the authors proposed a CNN that separates fat and water using real and imaginary data from six echo times of single-slice knee and head multiecho MRI. 

The conditional generative adversarial network (cGAN) architecture has been proposed and is widely used for image-to-image style transfer~\cite{isola2017image}, and GANs have been broadly applied in medical imaging for tasks such as data augmentation, image reconstruction and segmentation~\cite{yi2019generative,dar2019image}. A recent study has applied the cGAN architecture using six-channel 2D data~\cite{shen2020improved} to perform fat-water separation in addition to predicting the field map and R2* values. The authors showed that their method outperforms a standard U-Net. The work to date by the deep learning community shows that a variety of methodologies, mostly based on the U-Net, using 2D data work for several multiecho processing problems including fat-water separation and that when more echoes are used as input, the performance of deep learning models increases. Some authors have performed slice-by-slice predictions of 3D volumes but none of those found during the literature review have performed fat-water separation in fully volumetric data.

In this paper, we construct a deep-learning model that performs swap-free fat-water separation in two-point Dixon MRI using 3D data.  We formulate fat-water separation as a style transfer problem, where the fat and water volumes are predicted from the in-phase and opposed-phase volumes subject to constraints taken from MR physics of the Dixon technique, and compare it to conventional L1 loss. Our implementation of a cGAN works for neck-to-knee acquisitions using 3D~patches. We assess the model output quantitatively for data without fat-water swaps and visually for data with swaps, since the data are inherently wrong if the original data are affected by swaps and, thus, can not be used to compute reconstruction quality metrics. This study makes the following contributions:

\begin{itemize}
\item we have developed 3D cGAN models for fat-water separation,
\item we have demonstrated that single-channel and dual-channel input models are able to produce volumetric fat-water separation, with better results from dual-channel input,
\item we have shown through our evaluation that our model is fast and fixes all swaps without inducing false positives,
\item we have shown that our method is able to detect swaps via absolute differences with original data and it outperforms a closely-related method which consistently introduces a small number of false positives,
\item we have introduced of a new Dixon loss function that exploits the known model used for two-point Dixon fat water separation, and
\item we have shown that the Dixon loss potentially removes the need for ground truth labels.
\end{itemize}

\section{Materials \& Methods}

\subsection{Data}

In this study, we used the Dixon~MRI from the UK~Biobank abdominal imaging protocol. The Dixon acquisitions provided for each subject come in six separate series with four channels each: in-phase $(IP)$, opposed-phase $(OP)$, fat $(F)$ and water $(W)$. Two of the volumes are acquired at times of maximum (in-phase) and minimum (opposed-phase) difference and are used to derive water and fat volumes, where in theory $W=(IP+OP)/2$ and $F=(IP-OP)/2$. However, this ignores the contribution of field inhomogeneity on the phase for the complex-valued data acquired by the scanner $IP=(W+F)e^{-i\phi_0}$ and $OP=(W-F)e^{-i(\phi_0+\phi)}$, where the difference in phase is related to the difference in echo times (TE) and the static magnetic field $\phi=2\pi\psi\Delta\text{TE}$~\cite{ma2008dixon}. 

\begin{figure*}[!ht]
    \centering
    \includegraphics[width=\textwidth]{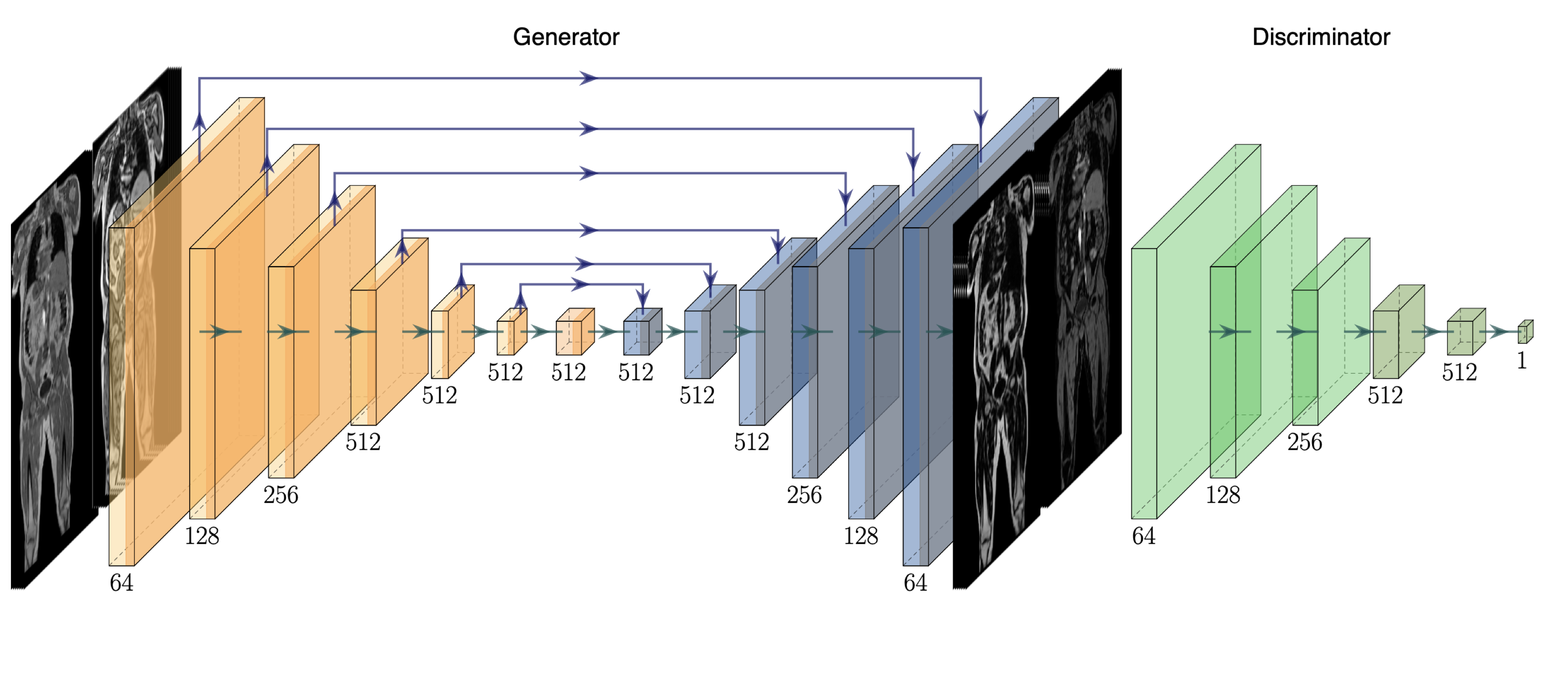}
    \caption{Proposed 3D cGAN architecture that takes $IP$ and $OP$ volumes as input and predicts the separated $\hat{F}$ and $\hat{W}$ volumes. Filter numbers for each convolutional block are shown below. The contracting path of the generator is in orange and the expansive path is in blue. The discriminator is shown in green, the different shades indicate the change in stride.
}
    \label{fig:cgan_architecture}
\end{figure*}

For the training data, fat-water separation was performed by the scanner reconstruction software on each Dixon series as provided by the UK~Biobank. We performed minor preprocessing to assemble the six series into a single volume for each channel~\cite{liu2021genetic} (Fig.~\ref{fig:data_and_swaps}~a-d). Briefly, the six series were resampled to the same resolution ($2.23 \times 2.23 \times 3.0~\text{mm}^3$), bias-field correction \cite{tustison2010n4itk} was performed on the in-phase volume and the estimated bias field applied to the other channels for each series, resulting in a final volume of size $(224, 174, 370)~\text{voxels}$. Bias-field correction was performed a second time on the blended in-phase volume and the estimated bias field was applied to the other channels. Estimating the bias-field is a particularly computationally expensive step, the entire process takes approximately 20~minutes per scan. We selected 1,027 participants as swap-free ground truth data for our experiments, where visual inspection of the fat and water channels was performed for each participant to ensure no substantial swaps were present. The participants were chosen to cover a broad range of age, gender and body mass index values. During the process of quality control, we also identified more than 70 participants with at least one fat-water swap in the original fat and water channels. We used those scans to verify swap correction by visual inspection. When developing the neural network model we used the $IP$ and $OP$ channels as our input data and the $F$ and $W$ channels as the training labels.

\noindent The North West Multicentre Research Ethics Committee (UK) approved the study and written informed consent was obtained from all subjects prior to the UK Biobank study entry.

\subsection{Neural Network Model}

The models for our fat-water separation experiments are based on the cGAN architecture~\cite{isola2017image}, originally developed for 2D image-to-image style transfer.  We define a generator $\mathcal{G}$ such that $\hat{F},\hat{W}=\mathcal{G}(IP,OP)$ is a direct mapping of the fat and water volumes from the in-phase (and opposed-phase) volumes, and a discriminator $\mathcal{D}$ that jointly determines whether or not the predicted volumes are truly fat and water or not. The objective function for the cGAN is given by
\begin{equation}\label{eqn:cgan_loss}
    L_\text{cGAN}(\mathcal{G},\mathcal{D}) =  E_{F,W}[\log\mathcal{D}({F},{W})]  + E_{IP,OP}[1 - \log\mathcal{D}(\mathcal{G}(IP, OP))],
\end{equation}
where the loss is driven solely by the performance of the discriminator. It has been shown that additional terms in the cGAN objective function may be added to ensure the predictions are similar to the ground truth, in our case $F$ and $W$ were used as ground truth labels in the L1 loss term \begin{equation}\label{eqn:l1_loss}
    L_1(\mathcal{G}) =  E_{F,W}[\|(F,W) - (\hat{F},\hat{W})\|_1],
\end{equation}
ensuring that each individual prediction matches the ground truth data. The solution 
\begin{equation}\label{eqn:solution}
    \overline{\mathcal{G}}=\arg\min_\mathcal{G}\max_\mathcal{D} L_\text{cGAN}(\mathcal{G},\mathcal{D}) + \lambda L_{1}(\mathcal{G})
\end{equation} 
is obtained by minimising the contribution of the generator to the objective function against a discriminator that tries to maximise it and the L1 term.


Fat and water volumes are predicted from the input volumes, illustrated in Fig.~\ref{fig:cgan_architecture}. The generator follows a U-Net architecture~\cite{ronneberger2015u} with six levels consisting of 3D convolutional layers (orange) with filter size~4 and stride~2, instead of pooling layers to move to the lower resolution levels. As the full volume dimensions were simply too large to be used as is, we set the input size to $(128, 128, 128)$ in order to overcome memory limitations. The number of filters are indicated at each level in Fig.~\ref{fig:cgan_architecture}. Up-sampling (blue) is performed via 3D transpose convolutional layers with filter size~4 and stride~2. The number of filters are indicated at each level. The discriminator network follows a sequential architecture using 3D convolutional layers with filter size~4 and stride~2 (green) except the last two layers with stride~1 (dark green), to adjust the PatchGAN discriminator network to assess the volumes on a patch size of $(16, 16, 16)$ voxels~\cite{isola2017image}. 

We created a data generator for the training process, which on every iteration selects a random participant and randomly crops a cube with $(128, 128, 128)$ voxels of the input data, as well as matching water and fat volumes for the output ground-truth labels. The data are jointly normalized to the 99th percentile of the maximum across the intensities of all channels to have all channels on the same scale between $0$ and $1$, and avoid spikes in signal intensity.

We chose the number of epochs and learning rate based on an initial fine-tuning experiment with 600 participants and 200 out-of-sample data. We chose the final parameters based on the results of the validation set. We found that the generator loss converged at $100$ epochs. The hyperparameter $\lambda$ was set to~$100$ following the methods of~\cite{isola2017image}. We set the learning rates to 0.0002, which also follows the methods of~\cite{isola2017image}. We performed a parameter sweep from 0.0001 to 0.01, with steps of 0.01 and 0.001, and found the best results and faster convergence to be with the original values. The batch size was limited by memory constraints, as we are working with large 3D arrays. We ran our experiments on a GeForce RTX~2080 Ti 12GB~GPU, code in (python~3.7.2, Keras~2.2.4~\cite{chollet2015keras} with tensorflow backend). We trained the model using the Adam optimizer and batch size~2 until convergence of the generator at $100$ epochs. Generator network weights of our best performing model to process a test subject of UK~Biobank Dixon MRI will be available upon publication at https://github.com/recoh/fat\_water\_separation.

\subsection{Experiments}

We trained three separate models with varying input data as well as incorporating a new loss function. Our first experiment utilised only the $IP$ channel as input to the cGAN and the $F$ and $W$ as ground-truth data to perform fat-water separation $IP \rightarrow \hat{F},\hat{W}$ using the L1~loss for the generator (Eqn.~\ref{eqn:l1_loss}) in a supervised manner. The second experiment utilised the dual-input model, using both the $IP$ and $OP$ channels as inputs, to perform fat-water separation $IP,OP \rightarrow \hat{F},\hat{W}$. For the third experiment, we propose a new Dixon loss function that exploits the physical model describing the relationship between the derived fat and water channels from the acquired $IP$ and $OP$ data in Dixon~MRI. This is a similar approach followed by~\cite{torop2020deep} for R2* estimation and~\cite{jafari2021deep}, where the authors incorporated physical models normally used to estimate final parameter maps in single-slice (2D) data. These approaches remove the dependency for ground truth labels and could have significant impact on problems where labeled data are difficult to obtain or where ambiguities in the data and/or model lead to corrupted predictions, such as fat-water swaps. Similar to the second experiment, this generator model used the $IP$ and $OP$ as model inputs to perform the fat-water separation $IP,OP \rightarrow \hat{F},\hat{W}$. We define $IP$ and $OP$ error terms using the $IP$ and $OP$ inputs as well as the $\hat{W}$ and $\hat{F}$ generator predictions, and combine them into a Dixon loss for the generator:

\begin{equation}
L_2(\mathcal{G}) = E_{IP} [||IP - (\hat{W} + \hat{F})||_2] + E_{OP} [||OP - |\hat{W} - \hat{F}|||_2].
\end{equation}

This Dixon loss function replaces the $L_1$ loss in (\ref{eqn:solution}).  We did not perform an equivalent fourth experiment with the Dixon loss for the single-input model generator. This is because there are obvious solutions when minimising for the in-phase channel on its own $IP = \hat{W} + \hat{F}$, and no other constraints being added to training. Obvious solutions where the relationship holds would be with either the $\hat{W}$ or $\hat{F}$ channels being empty, and the model output being equivalent to $IP$.

\begin{table*}[!tbp] 
\centering
\newcolumntype{C}{>{\centering\arraybackslash}X}
\caption{Quantitative assessment of fat and water predictions compared with the original data, using four-fold cross-validation on 800 scans. Values reported are average and standard deviation for the SSIM and PSNR values. The three models are: single-channel $IP\rightarrow\hat{F},\hat{W}$, dual-channel $IP,OP\rightarrow\hat{F},\hat{W}$ and dual-channel (Dixon generator loss) $IP,OP\rightarrow\hat{F},\hat{W}$.}
\label{tab:metrics}
\begin{tabular}{lcrrrr}
\addlinespace
& & \multicolumn{4}{c}{\makecell{Water} \makecell{Fat}} \\
\cmidrule(lr){3-6}
\multicolumn{1}{l}{Model} & \multicolumn{1}{c}{Run} & \multicolumn{1}{c}{SSIM} & \multicolumn{1}{c}{PSNR (dB)} & \multicolumn{1}{c}{SSIM} & \multicolumn{1}{c}{PSNR (dB)}
\tabularnewline
\cmidrule[\lightrulewidth](lr){1-6}\addlinespace[1ex]
$IP\rightarrow\hat{F},\hat{W}$ & 1 & 0.919~{\textpm}~0.011 & 24.28~{\textpm}~0.78 & 0.945~{\textpm}~0.009 & 24.70~{\textpm}~0.84 \tabularnewline
& 2 & 0.913~{\textpm}~0.012 & 24.07~{\textpm}~0.70 & 0.942~{\textpm}~0.008 & 24.45~{\textpm}~0.75 \tabularnewline
& 3 & 0.926~{\textpm}~0.009 & 24.74~{\textpm}~0.77 & 0.942~{\textpm}~0.008 & 25.07~{\textpm}~0.83 \tabularnewline
& 4 & 0.919~{\textpm}~0.010 & 24.35~{\textpm}~0.74 & 0.945~{\textpm}~0.010 & 24.55~{\textpm}~0.81 \tabularnewline
\cmidrule[\lightrulewidth](lr){1-6}\addlinespace[1ex]
$IP,OP\rightarrow\hat{F},\hat{W}$ & 1 & 0.961~{\textpm}~0.006 & 28.99~{\textpm}~0.91 & 0.975~{\textpm}~0.004 & 29.67~{\textpm}~0.96 \tabularnewline
& 2 & 0.962~{\textpm}~0.005 & 28.94~{\textpm}~0.82 & 0.972~{\textpm}~0.003 & 29.00~{\textpm}~0.80 \tabularnewline
& 3 & 0.966~{\textpm}~0.005 & 29.41~{\textpm}~0.83 & 0.976~{\textpm}~0.004 & 29.58~{\textpm}~0.85 \tabularnewline
& 4 & 0.963~{\textpm}~0.005 & 29.10~{\textpm}~0.84 & 0.974~{\textpm}~0.004 & 29.41~{\textpm}~0.83 \tabularnewline
\addlinespace
\cmidrule[\lightrulewidth](lr){1-6}\addlinespace[1ex]
$IP,OP\rightarrow\hat{F},\hat{W}$  & 1 & 0.930~{\textpm}~0.010 & 25.11~{\textpm}~0.83 & 0.953~{\textpm}~0.007 & 25.37~{\textpm}~0.91 \tabularnewline
(Dixon generator loss) & 2 & 0.928~{\textpm}~0.008 & 25.51~{\textpm}~0.85 & 0.949~{\textpm}~0.007 & 25.51~{\textpm}~0.85 \tabularnewline
& 3 & 0.935~{\textpm}~0.009 & 25.94~{\textpm}~0.91 & 0.952~{\textpm}~0.008 & 26.14~{\textpm}~0.96 \tabularnewline
& 4 & 0.924~{\textpm}~0.009 & 25.36~{\textpm}~0.89 & 0.951~{\textpm}~0.008 & 25.35~{\textpm}~0.88 \tabularnewline
\addlinespace
\bottomrule
\end{tabular}
\end{table*}

\begin{table*}[!tbp] 
\centering
\caption{Quantitative assessment of fat and water predictions compared with the original data, using all 800~scans from the cross-validation experiments for training and an out-of-sample test set of 227~scans for evaluation. Values reported are average and standard deviation for the SSIM and PSNR values.}
\label{tab:metrics_test}
\begin{tabular}{lcccc}
\addlinespace
 & \multicolumn{4}{c}{\makecell{Water} \makecell{Fat}} \\
 \cmidrule(lr){2-5}
Model & SSIM & PSNR (dB) & SSIM & PSNR (dB)
\tabularnewline
\cmidrule[\lightrulewidth](lr){1-5}\addlinespace[1ex]
$IP\rightarrow\hat{F},\hat{W}$ & 0.926~{\textpm}~0.010 & 24.77~{\textpm}~0.73 & 0.949~{\textpm}~0.008 & 25.22~{\textpm}~0.79 \tabularnewline
\cmidrule[\lightrulewidth](lr){1-5}\addlinespace[1ex]
$IP,OP\rightarrow\hat{F},\hat{W}$ & 0.967~{\textpm}~0.005 & 29.47~{\textpm}~0.86 & 0.977~{\textpm}~0.004 & 29.74~{\textpm}~0.91 \tabularnewline
\cmidrule[\lightrulewidth](lr){1-5}\addlinespace[1ex]
$IP,OP\rightarrow\hat{F},\hat{W}$  &0.939~{\textpm}~0.008&26.48~{\textpm}~0.90&0.953~{\textpm}~0.007&26.73~{\textpm}~0.92 \\
(Dixon generator loss) \tabularnewline
\addlinespace
\bottomrule
\end{tabular}
\end{table*}

\subsection{Evaluation}

To quantitatively evaluate the output of the models in our three experiments, we predicted the entire neck-to-knee $\hat{F}$ and $\hat{W}$ volumes by performing piece-wise predictions covering the entire input volumes. For each subject, we divided the $IP$ (and $OP$) into sixteen subvolumes of $(128, 128, 128)$ to match the model input size and reassembled the $\hat{F}$ and $\hat{W}$ predictions to get a full volume of $(224, 174, 370)$.

For all three experiments, we performed four-fold cross-validation using 800 images in a 75-25 split. We trained final models on the 800 subjects for each of the three experiments and evaluated them on the 227 subjects kept aside for testing. We used the structural similarity index measure (SSIM)~\cite{wang2004image} and Peak Signal-to-Noise Ratio (PSNR) to assess the predicted $\hat{F}$ and $\hat{W}$ volumes against the original data, as given by the internal scanner separation. The PSNR is given by
\begin{equation}
    \text{PSNR} = 10 \cdot \log_{10}\left(\frac{\text{MPI}^2}{\text{MSE}}\right),
\end{equation}
where MPI is the maximum pixel intensity and MSE is the mean square error. SSIM is defined via
\begin{equation}
    \text{SSIM}(x,y) = \frac{(2\mu_{x}\mu_{y} + C_{1})(2\sigma_{xy} + C_{2})}{(\mu_{x}^{2} + \mu_{y}^{2} + C_{1})(\sigma_{x}^{2} + \sigma_{y}^{2} + C_{2})},
\end{equation}
where $\mu_x$ and $\mu_y$ represent averages of the reference image windows $x$ and $y$, respectively, of the test images, $\sigma_x^2$ and $\sigma_y^2$ are the variances of the reference image windows and $\sigma_{xy}$ is their covariance. The terms $C_1$ and $C_2$ are dynamic range constants. SSIM values range from $-1$ to $+1$, the latter which is only achieved for two identical images. We computed the SSIM using the default sliding window size ($11 \times 11 \times 11)$ on the entire predicted $\hat{F}$ and $\hat{W}$ volumes. 

Due to the nature of swapped data, quantitative evaluation using image quality metrics is not possible. The original data that would be used as ground truth is inherently flawed and may not be used for this purpose. However, the absence of swaps in our predictions can be verified via visual inspection. We predicted the $\hat{F}$ and $\hat{W}$ volumes for more than 70~participants that were affected by swaps identified when selecting training datasets, and performed visual inspection on the volumes to identify whether or not the models were able to successfully perform fat-water separation where the scanner software failed. 

As a final qualitative evaluation of our model on swapped data, we used the predicted fat and water channels $(\hat{F},\hat{W})$ to perform segmentations using publicly available code~\cite{liu2021genetic,fitzpatrick2020large} in order to highlight the importance of correct fat-water separation. We implemented the \textit{dixonfix} method by Glocker et~al.~\cite{glocker2016correction} to compare our method with a closely-related alternative where the code is publicly available. In~\cite{glocker2016correction}, the authors trained the model using 23~subjects. Acknowledging the fact that we have more training data available, we trained the \textit{dixonfix} method using 150 fully assembled volumes of size $(224,174,370)$, a factor of 6.5~times more in terms absolute numbers of subjects used and possibly more in terms of array size. Instead of performing fat-water separation, \textit{dixonfix} predicts binary labels of swap locations via an intermediate step of fat-water separation using a regression forest. Swap labels are then computed using a graph-cuts algorithm. The labels are used to swap the fat and water data back into the original channel. 

Since our method directly predicts the fat and water channels, the two methods are fundamentally different. However, we are able to construct an analogous swap ``label map'' by thresholding the absolute differences of our $\hat{F}$ and $\hat{W}$ predictions with respect to the original fat and water data, where a swap in the original data will show up as a cluster of pixels with large difference in values. For this comparison, we used the final supervised dual-input model. To convert absolute differences into label maps, we used a threshold of $0.9$ and a minimum cluster size matching the smallest ones induced by \textit{dixonfix}. We then counted the number of false positive clusters for predictions of 50 out-of-sample participants (taken from the 200 kept aside for final testing), as well as absolute number of false positive voxels and the fraction of false positives compared to the entire volume.

\section{Results}

\begin{figure*}[!ht]
    \centering
    \includegraphics[width=\textwidth]{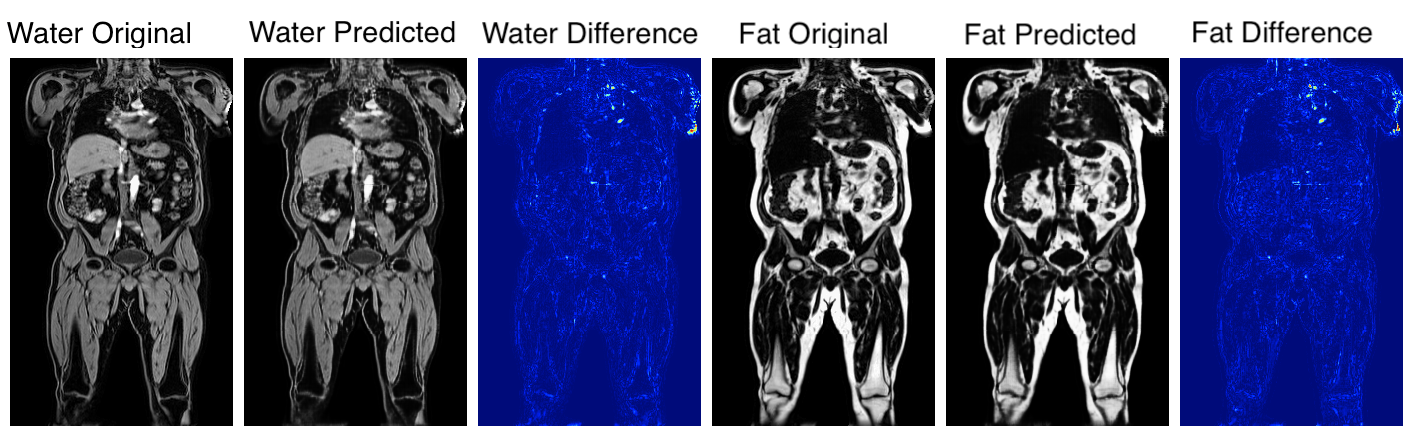}
    \caption{Original and predicted fat and water channels and their absolute differences. Predicted values were performed on a participant in the test set, where the model was trained on all 800~participants in the training set. Performance metrics calculated on the full 3D~volume are: $\text{SSIM}_W = 0.953$, $\text{SSIM}_F = 0.970$, $\text{PSNR}_W = 29.80\,\text{dB}$ and $\text{PSNR}_F = 29.85\,\text{dB}$.}
    \label{fig:fig3}
\end{figure*}

Quantitative evaluations of the three models, using four-fold cross validation on 800 participants, are provided in Table~\ref{tab:metrics}. The dual-channel supervised model is clearly superior to the single-channel supervised model, exhibiting higher SSIM and PSNR values for all runs in the cross-validation experiment.  The dual-channel model trained with generator Dixon loss performs better than the single-channel supervised model but worse than the dual-channel L1 generator model. This makes sense since the Dixon generator loss model does not benefit from ground truth labels, but has twice as much input information as the single-input model. Final models, trained on all 800 scans utilised in the cross-validation experiments, evaluated against an out-of-sample test set of 227~scans are shown in Table~\ref{tab:metrics_test}. 

\begin{figure*}[!ht]
    \centering
    \includegraphics[width=\textwidth]{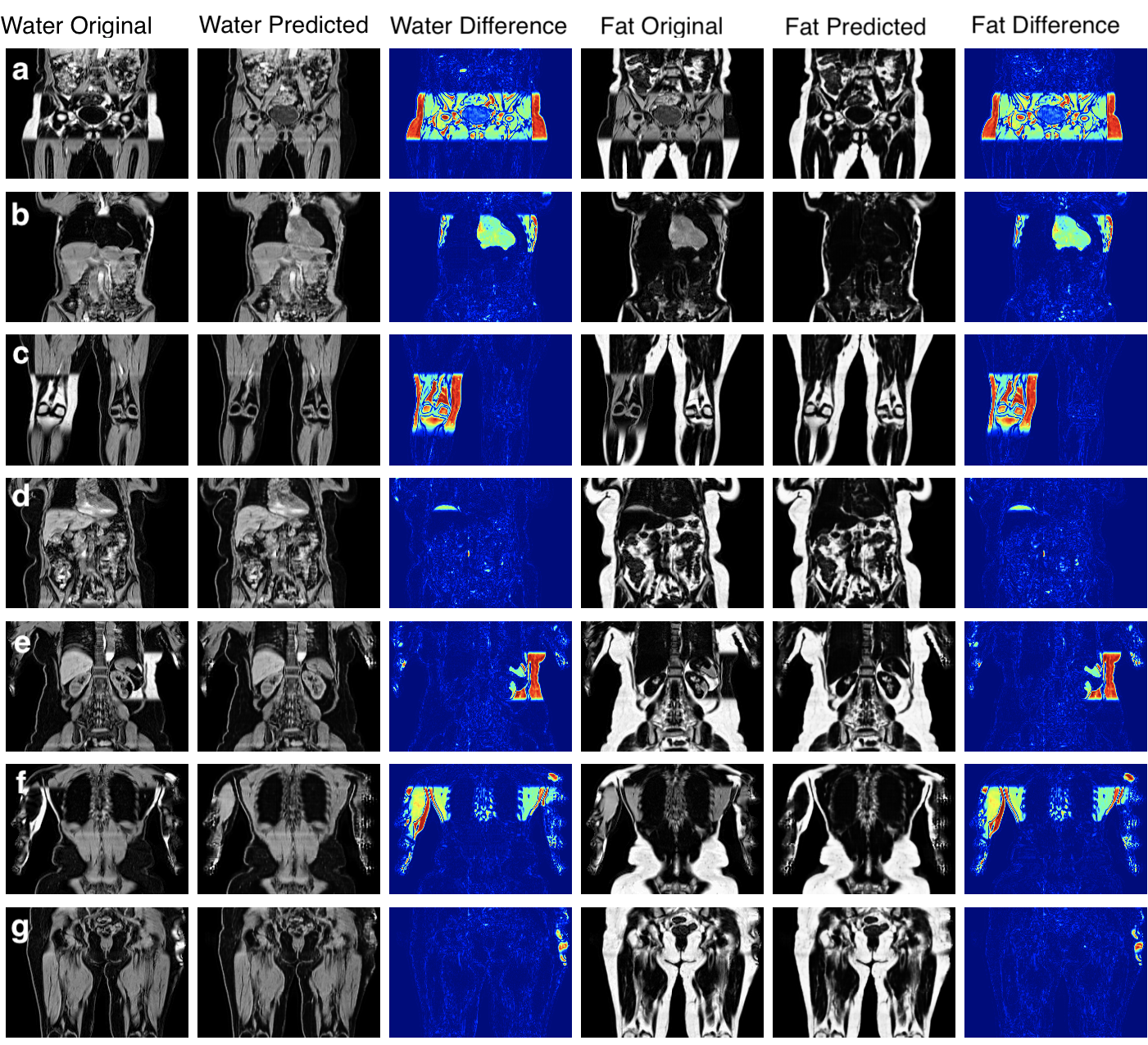}
    \caption{Model predictions of data where the original separated fat and water channels contain swaps. A full swap of the fourth series (a), a partial swap in second series (b), a swap in one leg (c), a swap at the top of the liver (d), a complex partial swap in the kidneys, spleen, and subcutaneous fat (e), a complex partial swap in the back and arm muscles at the edge of the field of view (f), and a partial swap at the extremities of the body due to inhomogeneities at the edge of the field of view (g).}    
    \label{fig:fig4}
\end{figure*}

\begin{figure*}[!ht]
    \centering
    \includegraphics[width=\textwidth]{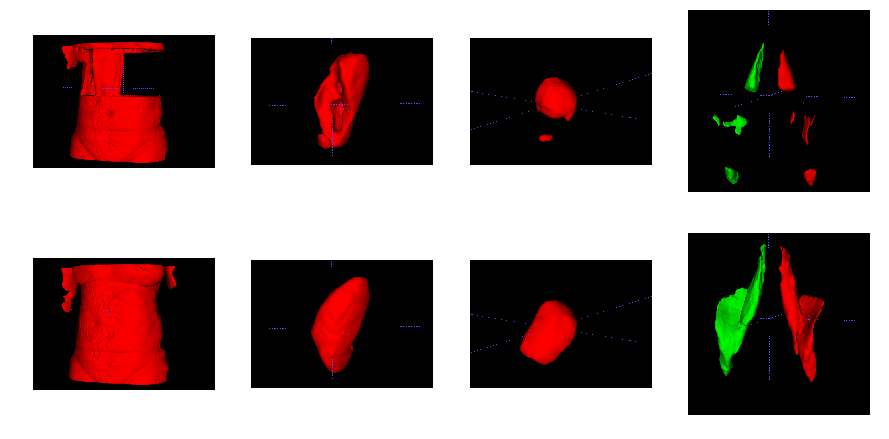}
    \caption{Impact of fat-water swaps on 3D segmentations of abdominal tissue and organs.  The top row shows segmentations generated using data that contained fat-water swaps, and the bottom row shows segmentations of the same tissue and organs using swap-free data from our model. From left to right: abdominal subcutaneous fat, left kidney, spleen, and left \& right iliopsoas muscles.}
    \label{fig:fig5}
\end{figure*}

Fig.~\ref{fig:fig3} shows the ground-truth data, predictions and their absolute difference for the fat and water channels of a participant in the testing set. Model predictions and the original data from the scanner for participants affected by various fat-water swaps are provided in Fig.~\ref{fig:fig4}. The examples were selected to illustrate the performance of our model in a variety of scenarios: data that are not affected by major swaps, data affected by swaps that cover an entire series in the acquisition (Fig.~\ref{fig:fig4}~a, c), as well as data displaying complex partial (Fig.~\ref{fig:fig4}~b, d, e) and/or boundary swaps (Fig.~\ref{fig:fig4}~f, g). The absolute difference images in columns~3 and~6 of Fig.~\ref{fig:fig4} highlight where the original data have been affected by a fat-water swap in the scanner reconstruction but the model correctly predicted the fat and water channels. Fig.~\ref{fig:fig5} provides examples of 3D segmentations using data that suffered from fat-water swaps (top row) and the segmentation when using our model predictions (bottom row) for the following organs and tissue (from left to right): abdominal subcutaneous adipose tissue, left kidney, spleen, and left/right iliopsoas muscles (red and green, respectively). The predictions shown in Figs.~\ref{fig:fig3} and~\ref{fig:fig4}, as well as the underlying volumes used for the 3D~segmentations in Fig.~\ref{fig:fig5}, are outputs of the final dual-input model, which performed the best across all of our experiments.

Our assessment of false positive rate between the \textit{dixonfix} method~\cite{glocker2016correction} and our method using 50 out-of-sample subjects that were free of swaps, has shown that our model induced zero false positives. On average, the \textit{dixonfix} method induced 4.29 swaps inside the body, with an average misclassification of $10,803 \pm 9,400$ voxels per subject, equivalent to an error rate of 0.074\% over the entire volume (including background voxels).

\section{Discussion}

It is still common for researchers to discard Dixon~MRI data due to the fact that fat-water swaps in the images reconstructed using the scanner software make them difficult to analyse. While these issues may make the data challenging, especially for non-imaging experts, we believe that the Dixon technique is a powerful and efficient tool for body composition studies and therefore warrants dedicated post-processing techniques to correct fat-water swaps. Our proposed model ensures that all data acquired in a study produces accurate quantitative results and that no resources, volunteers, patients, or user time are lost.

We have shown that our single- and dual-input models are able to predict swap-free fat and water volumes. Processing the entire neck-to-knee volumes, for example those found in the UK~Biobank abdominal protocol, takes approximately eight seconds per scan. We have established the high quality of our results through quantitative metrics such as PSNR and SSIM (with average values consistently $>0.95$) for the dual-input model in both cross-validation experiments (Table~\ref{tab:metrics}) and out-of-sample test data on final versions of the models (Table~\ref{tab:metrics_test}). We have shown qualitative performance on scans where the scanner software failed to adequately separate the fat and water channels during reconstruction in visual examples (Fig.~\ref{fig:fig3}) and by comparing to \textit{dixonfix} to establish our method does not induce fat-water swaps (false positives). As a final qualitative validation we used the predicted channels as input to 3D~segmentation models and have shown how the corrected data produces superior segmentations. 

We successfully separated fat and water channels for 3D Dixon MRI in our three experiments, where the best model utilises both the in-phase and out-of-phase data as input with L1 loss. We have shown that our method correctly separates $IP$ and $OP$ channels and is able to overcome a wide variety of fat-water swaps, including those explicitly excluded from the training data (e.g., swaps that completely cover one of the series acquired, cover half the series in the legs, top-of-the-liver swaps). However, minor fat-water swaps located at the boundary of the field of view or only involving the arms were included in the training data as they occur infrequently and at random anatomical locations.  We hypothesise that the infrequency and randomness of these types of swaps in the training data means that the model ignores them when optimising the generator. In Fig.~\ref{fig:fig4}, particularly rows e-g, it appears that the model is able to ignore and minimise the effect of the those swaps even though they were almost certainly present in the training data.  

When assessing the false positive rate of our model and comparing it to the \textit{dixonfix} method, no false positives were detected while the latter introduced on average an error of approximately $10,000$ incorrectly-swapped voxels per subject, corresponding to roughly $171\,\text{ml}$ in volume. This could have a significant impact on downstream analyses of structures such as visceral adipose tissue, muscles and in particular smaller abdominal organs (liver, spleen, kidneys, etc). While the authors in~\cite{glocker2016correction} pointed out that all swaps were corrected, they did underline the fact that every prediction induced false positives in the 3D volume.  These incorrectly-swapped voxels may appear in unimportant areas of the data, but we observed some induced swaps in structures such as skeletal muscle and adipose tissue. 
Depending on the downstream analyses performed after fat-water swap correction, such small errors may be acceptable. Further assessment on the impact of fat-water swap corrections with consistent false positives or false negatives (i.e., fat-water swaps that are ignored) in large-scale population studies is an interesting direction of investigation but outside the scope of this work. 

\section{Conclusion}
We have shown that our method allows for fast and reliable fat-water separation, the best results were produced by our dual-input model. We have also shown that our method correctly separates the fat and water channel where the scanner did not, without introducing false negatives. We highlighted the impact of our contribution by performing segmentation of original swapped data. By doing this, we also further displayed the high quality of the predictions. Since the CNNs were trained on original fat and water channels and produce good results with the model predictions, and CNNs are known for being highly specific and not able to generalise well. While the method is broadly applicable to two-point Dixon volumes, our model is intended for UK Biobank Dixon MRI. 

Future work will involve incorporating more of the preprocessing steps from our image analysis pipeline~(https://github.com/recoh/pipeline) into the neural-network model. For example, the input is assumed to have bias-field correction performed but this is computationally expensive. If our model assumes the input volumes (in-phase and opposed-phase) are not bias-field corrected, but trained using the bias-field corrected fat and water volumes, then the model will learn to correct the signal intensities in addition to accurate fat-water separation. This has the potential to decrease processing time by an order of magnitude. It is possible that the models already perform well on such unprocessed data and is something we will assess quantitatively going forward. The current version of our model does not work on the full volume, due to memory constraints. We believe the process would further benefit from working on the entire volume instead of the current patch-based implementation. Finally, we will explore the possibilities to turn our third experiment, using the Dixon generator loss function for the generator, into a fully self-supervised framework where the discriminator will only utilise the in-phase and opposed-phase channels as ground truth for the perceptive loss, completely removing the need for the fat and water ground-truth data in training. We also hope to be able to 
directly implement our method in the scanner reconstruction software, to help eliminate fat-water swaps from the source. 

\section{Acknowledgements}

We would like to thank Dr~Ben Glocker for his help in setting up the \textit{dixonfix} software.  

\bibliography{main}

\end{document}